\documentclass[twocolumn,prl]{revtex4}
\usepackage{graphicx}
\usepackage{subfigure}
\newcommand{\Ket}[1]{\vert \, #1 \, \rangle}

\begin{document}

\title{Resonant Cooling of Nuclear Spins in Quantum Dots}
\author{M. S. Rudner, L. S. Levitov}
\affiliation{
 Department of Physics,
 Massachusetts Institute of Technology, 77 Massachusetts Ave,
 Cambridge, MA 02139}


\begin{abstract}
We propose to use the spin-blockade regime in double quantum dots to reduce nuclear spin polarization fluctuations in analogy with optical Doppler cooling. 
The Overhauser shift brings electron levels in and out of resonance, creating feedback to suppress fluctuations.
Coupling 
to the disordered nuclear spin background is a major source of noise and dephasing in electron spin measurements in such systems.
Estimates indicate that a better than 10-fold reduction of fluctuations is possible.
\end{abstract}

\maketitle

A common goal across all areas of experimental physics is to reduce noise and fluctuations.
While this can often be achieved by cooling the entire system, in many situations only a few relevant degrees of freedom need to be ``cold.''
Directed cooling methods have been developed to specifically target these experimentally important degrees of freedom, 
allowing the relevant subsystem's 
temperature to be reduced by many orders of magnitude with little effect
on the rest of the system. 
One very important example of directed cooling that has led to great successes in the field of atomic physics\,\cite{Chu98} 
is the method of Doppler cooling\,\cite{Hansch75}, 
used to cool the velocity distribution of atomic and molecular gases.
Cooling is achieved by absorption of near-resonant monochromatic radiation,
with the velocity-dependent Doppler shift enhancing absorption 
for faster particles and suppressing it for slower particles.
In solid state systems, laser-like cooling was proposed for nanomechanical
resonators\,\cite{Hopkins03,Wilson-Rae04,Martin04}, and recently demonstrated for a superconducting qubit\,\cite{Valenzuela06}.


Advances in semiconductor technology have opened up many new possibilities for controlling microscopic degrees of freedom in solid state systems.
Experiments in the so-called spin-blockade regime 
of double quantum dots\,\cite{OnoTarucha} have demonstrated control over the charge and spin degrees of freedom of individual electrons, as well as control over the spin degrees of freedom of lattice nuclei due to their hyperfine coupling to electron spins\cite{Koppens, Petta}.
This same coupling of electron and nuclear spins also gives rise to the primary source of noise and dephasing in electron spin measurements due to the disordered nuclear spin background\cite{Petta,ESR}.
As a result, there is now a great impetus to better understand how to control and to cool the nuclear spin degrees of freedom in such systems\,\cite{Klauser06,Inarrea06}.

The key idea behind Doppler cooling is that the process that leads to cooling (absorption) receives feedback from the degree of freedom that is being cooled (velocity).
Here we propose an analogous scheme for cooling the nuclear spin polarization degree of freedom in a double quantum dot.
Using a simple model of spin-blockaded electron transport, we identify a window of detuning where nuclear spin evolution is controlled by the balance between competing rates of nearly resonant spin flip-flop processes in opposite directions.
These rates are sensitive to polarization via the Overhauser shift, which brings electron singlet and triplet states in-to and out-of resonance.
The interplay between opposing processes gives rise to a rich phase diagram that exhibits hysteretic behavior.
In direct analogy with Doppler cooling, we show that 
nuclear spin fluctuations are strongly suppressed in phases dominated by nearly resonant spin flips.
We quantitatively analyze this cooling effect and discuss possible
ways to optimize and employ it in experiments.


\begin{figure}
\includegraphics[width=3.4in]{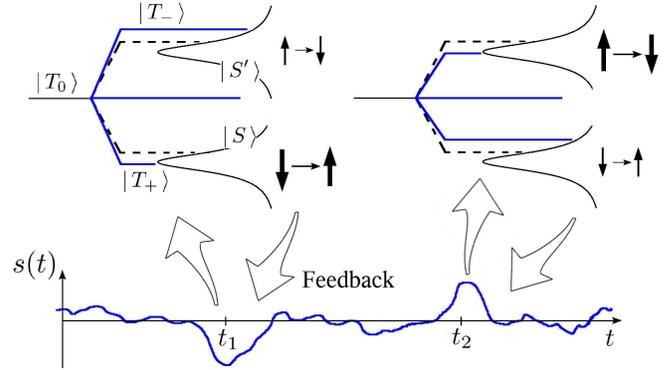}
\vspace{-4mm}
 \caption[]{Resonant suppression of nuclear spin fluctuations.
Net spin flip rates are peaked near the resonance of the electron triplet states with the broadened singlets $\Ket{S}$ and $\Ket{S'}$.
Dashed lines indicate the energies of the electron triplet levels $\Ket{T_\pm}$ in the steady state.
A fluctuation of polarization that 
changes the triplet level splitting leads to an imbalance of rates that drives polarization back to the steady state value. 
Fluctuations of either sign are suppressed (see times $t_{1,2}$).
}
\label{fig:Cartoon}
\vspace{-.25in}
\end{figure}

The key physical principle 
responsible for cooling in the spin-blockaded double dot is illustrated in Fig.\ref{fig:Cartoon}.
The two-electron triplet states $\Ket{T_\pm}$ are split by the Zeeman energy $E_Z = \pm g \mu_B [B +  \alpha\, s(t)]$, where $B$ is the applied magnetic field, and $\alpha s(t)$ is the Overhauser field felt by electrons due to hyperfine coupling to the nuclear spin system with instantaneous mean polarization $s(t)$.
The electron g-factor and hyperfine coupling in GaAs quantum dots
are $g_{\rm GaAs} \approx -0.4$ and $\alpha N  \approx -5\, {\rm T}$ where $N$ is the number of nuclear spins in the dot. 
Because $\alpha < 0$, nuclear spins polarized {\it against} the external field 
increase the total field seen by the electrons.


Transitions between electron triplet and singlet states occur due to the hyperfine interaction.
Each transition from $\Ket{T_-}$ to $\Ket{S'}$ is assisted by the flip of one nuclear spin from up to down.
Similarly, transitions from $\Ket{T_+}$ to $\Ket{S}$ are assisted by spin flips from down to up.
The rates of these transitions are greatly enhanced when electron singlet and triplet levels approach resonance.

In the steady state, the total rates of nuclear spin flips from down to up and from up to down are equal.
When this steady state is perturbed by a fluctuation of nuclear spin polarization in the direction opposite to the external field, 
electron Zeeman energy increases.
For electron states detuned as 
in Fig.\ref{fig:Cartoon}, this brings the levels $\Ket{T_+}$ and $\Ket{S}$ closer to resonance and pushes $\Ket{T_-}$ and $\Ket{S'}$ further out of resonance; the rate of nuclear spin flips into the direction of the external field becomes greater than the rate of opposing spin flips and polarization is driven back toward the steady state value.
Fluctuations in the opposite direction are suppressed analogously.
For materials where 
$\alpha$ is positive, cooling will be achieved for energy detuning
on the opposite side of the resonances where the sign of feedback is reversed.

To understand this cooling effect quantitatively and to determine its experimental signatures, we now present a detailed analysis.
We start by generalizing the model put forth in Refs.\cite{OnoTarucha,Baugh,Koppens,Petta} to describe spin blockade of their double quantum dot devices in terms of two-electron states $(0, 2)_s$, $(1, 1)_s$ and $(1, 1)_t$.
The states $(n_L, n_R)_{s/t}$ are labeled by electron occupation numbers $n_L$ and $n_R$ of the left and right dots.  
Subscripts denote the two-electron spin state, which can be singlet or triplet. 

Recent experiments in lateral double dots\,\cite{Koppens,Petta} were made in the regime where 
the singlet states $(0, 2)_s$ and $(1, 1)_s$ can be tuned through an avoided crossing.
Orbital hybridization makes the labeling scheme based on dot occupancy not suitable for the singlets.
Instead we will use the labels $S$ and $S'$ for singlets, and $T_{0,\pm}$ for triplets as shown in Fig.\ref{fig:Cartoon}.
Throughout this work we will take\footnote{
In previous work\,\cite{OneResonance} we discussed transport in 
the regime of large detuning, where the energy separation of the $(0, 2)_s$ and $(1, 1)_t$ states is much larger than 
that of the $(1, 1)_s$ and $(1, 1)_t$ states.
}
\begin{eqnarray}
E_{S',S} = {\textstyle\frac12}(\pm\sqrt{t^2 + \Delta^2} - \Delta)
,\ 
E_{T_0} = 0
,\ 
E_{T_\pm} = \pm E_Z
,
\end{eqnarray}
where $t$ is the tunnel splitting, and
$\Delta$ is the detuning from the singlet energy crossing.


Transport occurs as a sequence of tunneling events.
Charge moves from the source to the left dot, through the right dot, and finally to the drain.
Starting with one electron localized on the right dot, 
the second electron tunnels into the left dot to form one of the five two-electron states $\{S, S', T_0, T_\pm\}$.
The probabilities for filling these states are determined by the relative rates of tunneling in from the source lead.
Because the three triplets have the same $(1, 1)$ orbital wavefunctions, the rates for tunneling into any of these states are the same.


The mechanism of charge transfer to the drain lead is very different for the singlet and triplet states\,\cite{OnoTarucha},
as the coupling of the $(0, 2)$ state to the drain is much stronger than that of the $(1, 1)$ states. 
Charge moves easily to the drain from the states $S$ or $S'$, since they both have a finite component in the $(0, 2)_s$ state.
When one of the triplet states is filled, however, the out-flow of charge is blocked; Pauli's Exclusion Principle and the high orbital excitation energy prevent the system from forming the $(0, 2)_t$ state.
Residual current is due to hyperfine flip-flop scattering into the singlet states $S$/$S'$, exchange with the leads, or 
tunneling through virtual excited states.

The rate of spin-flips due to hyperfine coupling is governed by the near resonance of 
the triplet states $T_{\pm}$ with the singlet states $S$ and $S'$. 
We describe the nuclear subsystem by the populations $N_{\pm}$ of the up and down nuclear spin states, neglecting their spatial variation.
The energy-dependent hyperfine spin flip transition rates $W_{\pm}^{{\rm HF}}$ are the sum of rates for transitions to $S$ and $S'$, calculated using Fermi's Golden Rule:
\begin{eqnarray}
 \label{eqnFGR}
 W_{\pm}^{{\rm HF}} = \frac{2 \pi}{\hbar} \alpha^2 N_\mp \left[\frac{A \gamma}{\varepsilon_\pm^2 + \gamma^2} + \frac{A' \gamma'}{(\varepsilon'_{\pm})^2 + (\gamma')^2}\right] 
\end{eqnarray}
where $A$ is the orbital overlap factor between the triplet state $T_\pm$ and the singlet $S$, $\varepsilon_{\pm}$ is the energy difference between $T_\pm$ and $S$, and $\gamma$ is the width of $S$.
We assume a Lorentzian lineshape to allow explicit calculation.
Primed terms represent the same quantities, calculated with respect to singlet state $S'$.
Throughout this paper we take $A = A' = 1$ without loss of generality.

The net spin flip rates are controlled by the competition between the rates of hyperfine spin flips and non-spin-flip tunneling processes in each channel\,\cite{OneResonance}.
Assuming a single energy independent non-spin-flip escape rate $W^{\rm in}$ and equal tunneling-in probabilities of 1/4 for each of the triplets, the net nuclear spin flip rates are given by
\begin{eqnarray}
  \label{eqnPumpRates}
   \Gamma_{\pm} = \frac{W_{\pm}^{{\rm HF}}}{W_{\pm}^{{\rm HF}} + W^{{\rm in}}}\frac{I_0}{4}
,\quad
I_0 = \frac4{\tau_- + \tau_0 + \tau_+}
.
\end{eqnarray}
where $I_0$ is the inverse of the average dwell time in $T_{0, \pm}$ with $\tau_0^{-1} = W^{\rm in}$ and $\tau_\pm^{-1} = (W^{\rm in} + W_\pm^{\rm HF})$.
We assume that the time scales for loading and for decay of the singlets are much shorter than the average triplet lifetime, and that the $T_0$ state does not decay by hyperfine transitions due to lack of energy conservation.
The spin flip rates $\Gamma_\pm$ receive feedback from nuclear polarization through the combination of the resonant energy dependence (\ref{eqnFGR}) and the Overhauser shift of electron levels\,\cite{OneResonance}.

We now study the flow of the average nuclear polarization $s \equiv N_+ - N_-$.
Because every spin flip changes $s$ by two, polarization evolves according to 
\begin{eqnarray}
\label{eq:V(s)}
\dot{s} = V(s) \equiv 2\left(\Gamma_+ - \Gamma_-\right) - \Gamma_{\rm rel} s,
\end{eqnarray}
where $\Gamma_{\rm rel}$ is a phenomenological relaxation rate due to nuclear spin diffusion and thermalization.
In the absence of a non-spin-flip escape path, $W^{\rm in}=0$ and every spin that enters the dot must flip its spin; $\Gamma_+ = \Gamma_-$ and no spin can be pumped.
When $W^{\rm in}$ is finite, a net polarization rate $V(s) \neq 0$ is achieved by tuning near one of the singlet-triplet resonances where spin flip processes are favored in one channel while non-spin-flip tunneling is favored in the other.
Polarization flows toward fixed points $V(s^*) = 0$ which give the steady state values of nuclear polarization.

\begin{figure}
\includegraphics[width=3.4in]{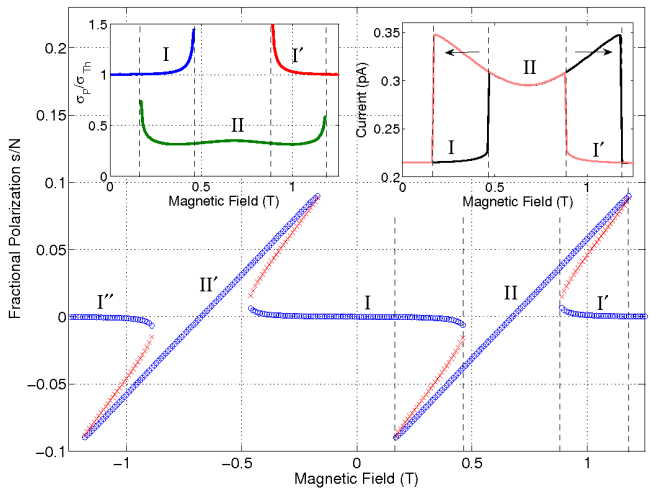}
\vspace{-3.5mm}
\caption[]{Steady state polarization $V(s^*) = 0$ versus magnetic field found numerically using (\ref{eq:V(s)}), (\ref{eqnPumpRates}), and (\ref{eqnFGR}) with $N = 10^7$, $t = 35 \,\mu {\rm eV}$, $\Delta = 3 \,\mu {\rm eV}$, $W^{\rm in} = 10^6\, s^{-1}$, and $\Gamma_{\rm rel} = 0.5 \,s^{-1}$.
Blue circles denote stable fixed points $V'(s^*; B) < 0$, while red crosses denote unstable fixed points $V'(s^*; B) > 0$.
Phases of type I are relaxation dominated, while phases of type II are resonance dominated.
Left Inset: Width of the nuclear spin distribution relative to the thermal width $\sigma_{\rm Th} = \sqrt{N}$.
Right Inset: Hysteretic current with magnetic field sweep,
calculated using $I_0$ defined in (\ref{eqnPumpRates})
(see discussion before Eq.(\ref{eq:I_full})).
Arrows indicate sweep direction.}  
\label{fig:PhaseDiagram}
\end{figure}

With the help of the stability condition, $V'(s^*) < 0$, we find that $V(s)$ contains 1, 2, or 3 stable fixed points 
depending on the values of parameters such as detuning, magnetic field, and relaxation rate.
In Fig.\ref{fig:PhaseDiagram} we plot $s^*$ as a function of magnetic field at constant detuning with realistic parameter values\,\cite{Koppens,Petta}.

Examining Fig.\ref{fig:PhaseDiagram}, we identify two distinct types of phases,
unpolarized (labeled I) and polarized (labeled II).
Phases of type I are relaxation dominated, and are characterized by nearly zero average polarization.
Spin flip transitions are off-resonant in these phases, and nuclear spin dynamics are dominated by noise and relaxation.
Phases of type II are resonance or polarization dominated.
In these phases, spin flip transitions are resonant in both directions, allowing for the build up of nuclear spin polarization and resonant suppression of fluctuations.
The existence of such phases is made possible by the close proximity of both resonances $S$ and $S'$.

With moderate relaxation $\Gamma_{\rm rel}$, we find that the unpolarized state is stable at zero applied magnetic field (phase ${\rm I}$).
This phase remains stable as magnetic field is increased, until a bifurcation point is reached (approximately $B = 0.5 \ {\rm T}$ for the case shown).
At this point, the Zeeman-split triplet levels reach far enough into the tails of the singlet resonances $S$ and $S'$ 
for the hyperfine spin-flips to overcome relaxation and trigger an instability towards a spin-polarized state (phase II).

After entering phase II, polarization changes smoothly from negative to positive values as $B$
is increased.
At even higher fields, a second bifurcation point is reached where the polarized phase terminates ($B \approx 1.2\ {\rm T}$ in the example shown).
Polarization flows back to 0 and the system returns to the relaxation-dominated phase ${\rm I}^{\prime}$.
On sweeping back towards lower fields, the system follows a different path.
This hysteresis is manifested in a double step of current for the magnetic field sweep described above (see Fig.\ref{fig:PhaseDiagram} inset), which is in striking resemblance to the current instabilities observed in Ref.\cite{OnoTarucha}.

To estimate nuclear spin fluctuations in the phases I and II, we shall analyze the behavior of the probability distribution $\rho(s, t)$.
Sequential electron transport through the spin-blockaded double dot is accompanied by a stochastic series of nuclear spins flips that change the net polarization in steps of $\pm 2$.
The polarization variable $s$ executes a directed random walk on the interval $[-N, N]$, where $N \approx 10^6 - 10^7$.
We describe this stochastic dynamics by the Fokker-Planck equation \cite{VanKampen}:
\begin{eqnarray}
  \label{eqnFP}
  \frac{\partial}{\partial t} \rho(s, t) = 
\frac{\partial}{\partial s}\left(D\, \frac{\partial}{\partial s}\rho(s, t) - V\,\rho(s, t) \right)
\end{eqnarray}
where $V(s) \Delta t = \langle \Delta s \rangle$ and $D(s) \Delta t = 1/2 \langle \Delta s^2 \rangle$,
with $V(s) = 2\left(\Gamma_+ - \Gamma_-\right)$, and $D(s) = 2\left(\Gamma_+ + \Gamma_-\right)$.
Here $\Delta t$ is a short interval on the order of one electron transit time.

In the simplest model of relaxation, polarization decays as the result of random flipping noise: with microscopic rate $\Gamma_{\rm rel}$, a random spin is selected and its z-component flipped.
The spin-flip rates (\ref{eqnPumpRates}) are then changed to
\begin{eqnarray}
\tilde\Gamma_\pm=\Gamma_\pm + \Gamma_{\rm rel}
N_{\mp}/N
,\quad
N_\pm={\textstyle \frac12} (N\pm s).
  \label{eqnThermalNoise}
\end{eqnarray}
This model reproduces the phenomenological relaxation term $-\Gamma_{\rm rel}s$ used in Eq.(\ref{eq:V(s)}) and accounts for broadening of the distribution due to coupling to the thermal bath.

The steady state of (\ref{eqnFP}) is
of the form $\exp \int_{s_0}^{s} V(s')/D(s')  ds'$.
One can easily check that this distribution is peaked around the stable fixed points $V(s^*) = 0,\ V'(s^*) < 0$ found earlier.
When fluctuations of $s$ about $s^*$ are small (i.e. away from bifurcation points),
we linearize $V/D$
and obtain a Gaussian steady state 
distribution peaked at $s = s^*$, with 
rms width $\sigma_P = \sqrt{D(s^*)/V'(s^*)}$.

In the absence of external driving, the thermal noise and nuclear spin diffusion responsible for $\Gamma_{\rm rel}$ cause the system to relax to the fully random high temperature distribution with zero mean $\langle s \rangle = 0$ and 
rms width $\sigma_{\rm Th} \equiv \sqrt{\langle s^2 \rangle} = \sqrt{N}$
\footnote{
This result is true for 1D (Ising) spins.  
The width of the disordered state of 3D spins is $\sigma_{\rm Th} = \sqrt{N/3}$.
}.
As current is restored and as the spin flip processes are brought closer into resonance, the net suppression of fluctuations is determined by a competition between confinement due to the resonant spin flip dynamics and broadening due to 
$\Gamma_{\rm rel}$.

\begin{figure}
\includegraphics[width=3.375in]{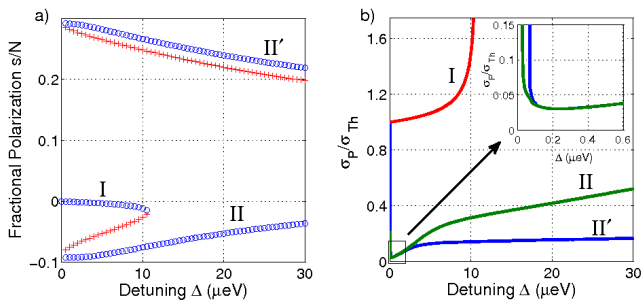}
 \caption[]{Polarization steady states (a) and widths (b) 
{\it vs.}  detuning.
Notation is the same as in Fig.\ref{fig:PhaseDiagram}.
Cooling is most efficient near $\Delta \approx \gamma$, where both resonances strongly contribute to the suppression of fluctuations.
Parameters: $N = 10^7$, $W^{\rm in} = 10^8\, s^{-1}$, $t = 50\, \mu{\rm eV}$, $\gamma = \gamma' = 0.1\, \mu{\rm eV}$, and $B = 0.5\, T$.
}
\label{fig:WidthPlots}
\end{figure}

The ratio $\sigma_P/\sigma_{\rm Th}$ serves as a quantitative measure of the strength of cooling.
As Fig.\ref{fig:PhaseDiagram} illustrates, phases of type ${\rm I}$ are characterized by $\sigma_P/\sigma_{\rm Th} \ge 1$ and thus do not exhibit cooling.
The polarized phase II, however, displays a significant suppression of fluctuations.  
Away from the edge of the instability, the strength of cooling is insensitive to magnetic field.

In contrast, the efficiency of cooling depends very strongly on detuning.
In Fig.\ref{fig:WidthPlots} we plot polarization and $\sigma_P/\sigma_{\rm Th}$ as a function of detuning in each of the three phases I, II, and ${\rm II}^\prime$ present for the chosen parameters.
Maximal cooling occurs at small positive detuning
\begin{eqnarray}
\Delta_\ast \approx \gamma,
\end{eqnarray}
where the system can benefit from the resonant character of spin flip rates for both the $T_+$ and $T_-$ states simultaneously (see Figs. \ref{fig:Cartoon} and \ref{fig:WidthPlots}b).
With both resonances $S$ and $S'$ contributing to the Fokker-Planck effective restoring potential, a greater than 10-fold suppression of thermal fluctuations can be achieved (see Fig.\ref{fig:WidthPlots}b inset).

\begin{figure}
\includegraphics[width=2.8in]{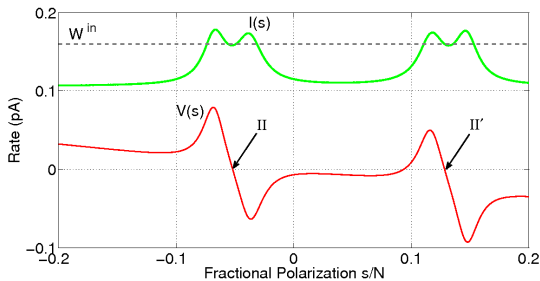}
\vspace{-3.3mm}
\caption[]{Current $I(s)$, Eq.(\ref{eq:I_full}), as a diagnostic of nuclear polarization. Results for $\beta=0$ are shown, 
when the non-spin-flip escape rate $W^{\rm in}$ is due to tunneling
back to source lead.
Note the non-monotonic dependence of current on polarization.
Parameters: $N = 10^7$, $t = 24 \,\mu {\rm eV}$, $\Delta = 4 \,\mu {\rm eV}$, $B = 0.2\, T$, $W^{\rm in} = 10^6 s^{-1}$, and $\Gamma_{\rm rel} = 0.1 \,s^{-1}$.
}
\label{fig:Rates}
\end{figure}

How are these phenomena reflected in device current?
Depending on the mechanism(s) responsible for the non-spin-flip escape rate $W^{\rm in}$, the current $I$ transmitted through the device may differ from the rate $I_0$ in (\ref{eqnPumpRates}).
Only electrons that move from source to drain count toward the flow of current; those that jump back to the source do not contribute.
Thus we introduce a parameter 
$0\le\beta\le1$ that specifies the fraction of $W^{\rm in}$ due to non-spin-flip escape to the drain, yielding
\begin{eqnarray}
\label{eq:I_full}
  I = \frac{I_0}{4}\left(1 + \beta + \frac{\beta W^{\rm in} + W_+^{\rm HF}}{W^{\rm in} + W_+^{\rm HF}} + \frac{\beta W^{\rm in} + W_-^{\rm HF}}{W^{\rm in} + W_-^{\rm HF}}\right).
\end{eqnarray}
When $\beta = 1$, all electrons 
exit to the drain and $I = I_0$.
We stress that the spin flip dynamics depend only on $I_0$ and not on the mechanisms responsible for $W^{\rm in}$ or on $\beta$.
Figure \ref{fig:Rates} illustrates that current (\ref{eq:I_full}) has a complicated non-monotonic dependence on $s$, but nonetheless can be used as an indicator of polarization.



We have shown that resonant hyperfine spin flips can be used to significantly suppress nuclear spin fluctuations in spin-blockaded double quantum dots
by a mechanism analogous to Doppler cooling.
Optimal cooling occurs at small positive detuning $\Delta_* \approx \gamma$ where both triplet-singlet resonances provide strong feedback through the Overhauser shift.
The suppression of nuclear spin fluctuations 
will manifest itself dramatically in electron spin resonance (ESR) measurements\,\cite{ESR}.
ESR experiments will both benefit from cooling through increased dephasing times and reduced fluctuations in Zeeman energy, and serve as a powerful way to detect it.

We thank L. M. K. Vandersypen and F. H. L. Koppens for useful discussions,
and acknowledge support (MR) of the DOE CSGF fellowship, grant DE-FG02-97ER25307.

\vspace{-6mm}


\end{document}